\documentclass[onecolumn,preprintnumbers,amsmath,amssymb,floatfix]{revtex4}

\usepackage{graphicx}

\usepackage{bm}
\usepackage{amsfonts}
\usepackage{lineno,hyperref}
\usepackage{array}
\usepackage{microtype}
\usepackage{float}
\usepackage{epstopdf}
\usepackage{mathrsfs}
\usepackage{color}

\begin{document}
	
	
	
	

	\title{Co-existence of alternative Generalized Chaplygin Gas and other Dark Energies
		in the Framework of Fractal Universe}

	\author{Mahasweta Biswas}
	\email{mahasweta.bsc11@gmail.com}
	\address{Department of Mathematics, Sonarpur Mahavidyalaya, South 24 parganas, Kolkata-700 149, India}
	
	\author{Rikpratik Sengupta}
	\email{rikpratik.sengupta@gmail.com}
	\address{Department of Physics, Aliah University, Action Area II, Newtown, Kolkata 700156, West Bengal, India}
	
	\author{Ujjal Debnath}
	\email{ujjaldebnath@gmail.com}
	\affiliation{Department of
		Mathematics, Indian Institute of Engineering Science and
		Technology, Shibpur, Howrah-711 103, India.}
	
	\begin{abstract}
	 We have explored the possibility of the co-existence of two forms of dark energy in the form of an alternative Generalized Chaplygin Gas (GCG) along with a scalar field of field theoretic or extra dimensional origin in the background of a fractal universe. The essential physical model parameters have been computed and their variations have been plotted. Fractal cosmology in an universe dominated by the alternative GCG (overcoming the drawbacks of the conventional form of GCG) has also  been studied by computing the equation of state, deceleration and statefinder parameters. Their variations with the redshift have also been studied. We have tried to construct alternative cosmological models deviating from standard cosmology that can be put to observational testing.
	\end{abstract}


\maketitle

\section{\normalsize\bf{Introduction}}

The idea of fractal cosmology or fractal universe (FU) can be approached from a number of completely different directions. The central idea revolves around the viability of the large scale homogeneity underlying the Cosmological Principle in the standard scenario. As the distribution of galaxies have been studied on larger scales with improved observational infrastructures over the last few decades as opposed to previous observations focusing on angular coordinates, the presence of both voids and structures in the large scale spatial distributions have been noted, which makes the assumption of homogeneity debatable \cite{Davis,Rees,Joyce}. This has encouraged the idea of a conditional Cosmological principle \cite{M1,M2,Coleman}. A fractal is a non-analytical structure characterized by fundamentally altered distribution of statistically identical points at all scales that is supposed to replicate the large scale distribution of matter in the universe \cite{Joyce2}. Another inconsistency of the standard scenario with the observed large scale cosmology is the inability to explain the current accelerating phase of the universe \cite{Riess,Perlmutter}. This can be achieved by modifying either the matter or the geometry sector (or both) of the Einstein Field Equations (EFE). The modification to the matter sector involving exotic components violating the energy conditions shall be discussed later in more details. The possible geometric modifications leading to an accelerating universe are diverse in nature ranging from reconstruction of the Einstein-Hilbert action to involve effective higher dimensional effects including an induced quantum back-reaction arising from braneworld gravity theories \cite{Sahni,DGP} to modifying the Lagrangian to include complicated higher order functions of the scalar curvature \cite{fr} or both curvature and trace of the stress-energy tensor \cite{frt}. A novel way of modifying the universal dynamics to incorporate the late-time accelerating phase also involves accounting for spatial inhomogeneities \cite{TB1,TB2,TB3}, from where echoes of a FU appear from a different view of large scale infra-red (IR) modification to standard cosmology. Also, the fractal property of the universe is exhibited in Linde's chaotic inflationary scenario on large scales where the bubble universes are formed by the inflating spacetime due to nucleation points arising from peaks created by the evolving scalar inflaton field \cite{Linde}.

Again, the idea of fractal cosmology finds relevance from certain aspects of quantum effects in the small-scale physics of high energy ultra-violet (UV) corrections to the standard gravitational scenario. Besides the two mainstream candidates for quantum gravity, namely Super-string/M-theory and Loop Quantum Gravity (LQG), there are some alternative approaches like causal dynamical triangulation\cite{cdt}, asymptotically safe gravity \cite{as1,as2} and Horava gravity \cite{h1,h2} which suggest spacetime to possess a fractal structure due to an evolving dimensionality of space with time, where at early times when UV corrections are predominant, the dimensionality is less than four to improve the renormalizability and usual four spacetime dimensions are recovered when UV correction effects loose their significance at lower energy scales. There is an isotropic scaling of the space and time coordinates and the gravitational action is modified due to the replacement of the standard measure in the action by a non-trivial measure, as we shall see in the following section. As already mentioned, the late-time acceleration of the universe can be accounted for by using exotic matter that violates the energy conditions. From the  EFE, we know that to account for cosmic acceleration it is necessary that the Strong Energy Condition (SEC) be violated. Observations from the Cosmic Microwave Background (CMB), in combination with large-scale structure surveys, provide increasingly precise constraints on the equation of state (EoS) parameter of dark energy (DE), $\omega$. Earlier analyses, such as those from the Two-Degree Field Galaxy Redshift Survey (2dFGRS) and CMB data, reported constraints of $-1.61 < \omega < -0.78$ at the 95 percent confidence level \cite{Percival2002}. However, more recent and precise measurements from the Planck 2018 CMB analysis \cite{Planck2020} and the extended Baryon Oscillation Spectroscopic Survey (eBOSS) of the Sloan Digital Sky Survey (SDSS) \cite{Alam2021} place much tighter bounds on $\omega$. The latest results indicate that for a flat $\Lambda$CDM model, the DE EoS is constrained to $\omega = -1.03 \pm 0.03$ (Planck 2018, TT,TE,EE+lowE+lensing+ Baryon acoustic oscillations (BAO)) and $\omega = -1.018 \pm 0.032$ (SDSS eBOSS 2020, combining BAO, type Ia supernovae (SNe Ia), and CMB data) at the 95 percent confidence level. These findings are more or less consistent with a cosmological constant ($\omega = -1$) but enforces a minimal deviation from the standard $\Lambda$CDM model. From all of this observational data also suggests that a better fit is provided for a super-negative EoS, such that the EoS parameter $\omega<-1$ which implies violation of the Null Energy Condition (NEC) \cite{E1}-\cite{E3}. In these papers it has been shown that a better fit can be provided to the observed cosmological parameters if the DE EoS crosses the phantom divide. In this paper, we will be concerned with a number of such exotic matter candidates and study their viability in the background of FU. \\

The accelerated expansion of the universe can be explained through the dynamics of scalar fields with non-standard kinetic terms. This framework suggests that a single scalar field can drive both inflation in the early universe and act as DE in its later stages. The unique properties of Dirac-Born-Infeld (DBI) actions introduce non-linear effects that could have observable consequences in the CMB and large-scale structures, while also providing a geometric interpretation linked to brane-world scenarios in string theory. Thus, DBI models offer a compelling connection between inflationary dynamics and DE, enriching our understanding of cosmic evolution \cite{DBIDarkEnergy2012}, \cite{DBIScalarFields2007}.
Tachyon DE is a theoretical model in cosmology that uses tachyonic fields—associated with hypothetical faster-than-light particles—to explain DE, which drives the universe's accelerated expansion \cite{InteractingTachyon2012}, \cite{AgegraphicTachyon2009}. These models feature a non-standard kinetic term and a potential energy that allows the tachyon field to evolve dynamically, transitioning the universe from deceleration to acceleration. This framework provides insights into the nature of DE and its role in cosmic evolution, offering unique predictions for structure formation and the fate of the universe. Dilation DE is a cosmological concept that connects DE dynamics to scalar fields undergoing dilation or scaling transformations \cite{TwoFieldDilaton2009},\cite{DilatonDarkEnergy2002}. In this model, the scalar field's energy density adapts to the universe's expansion, potentially providing a stable form of DE. This approach helps address the fine-tuning and coincidence problems by allowing the energy density to evolve over time, offering insights into the accelerated expansion of the universe and its fundamental forces.\\

The quintessence \cite{q1,q2} scalar field models provide the standard framework for DE models, where the scalar field is assumed to have a steep potential $V(\phi)$ that satisfies the criterion $\frac{V''V}{V'^2}\geq1$, prime ($'$) denoting derivative with respect to the scalar field $\phi(t)$. The advantage of having such a potential is that the field can evolve to an identical path from a large class of initial conditions. These models are often dubbed as tracker models as the energy density and equivalently the EoS of the scalar field remains in close approximity to that of the dominant background matter for a large part of the cosmological evolution. A modified form of the quintessence was soon proposed \cite{k1,k2} known as the k-essence where the kinetic term in the Lagrangian is non-canonical. Since the solution here is that of a dynamical attractor, the $\Lambda$-like behaviour persists only at the onset of matter domination. The dynamical attractor property also ensures that the initial conditions do not influence the late-time behaviour. If the EoS parameter of the DE component crosses the divide $\omega=-1$ such that the NEC is violated, then it is called the phantom DE \cite{Pe1,Pe3}. However, this results in problems like quantum instabilities of the vacuum as the kinetic term in the effective scalar field model has a negative sign \cite{Moore}. Hessence \cite{wei} is another form of DE with a non-canonical scalar field that is complex in nature possessing an internal degree of freedom and capable of crossing the phantom divide without instabilities. For an appropriate choice of the hessence potential, a Chaplygin gas like behaviour can be reproduced at late times. Also, Copeland et al. \cite{Copeland} approach to understanding the universe's accelerating expansion, focusing on DE. One discussed observational evidence, various models (like quintessence and tachyon), and the role of cosmological scaling solutions. Additionally, it explores tracking solutions, potential future singularities, and modifications to gravity that could account for late-time acceleration without introducing new DE.

The interest in exploring dark energy (DE) models, such as the Chaplygin gas, stems from the persistent theoretical issues with the cosmological constant $\Lambda$, despite its success in matching observations. The cosmological constant is plagued by the well-known fine-tuning and coincidence problems \cite{weinberg1989}, prompting cosmologists to investigate dynamic alternatives. These include models like quintessence \cite{q1}, k-essence \cite{armendariz2001}, and unified theories such as the Chaplygin gas \cite{kamenshchik2001, bento2002}. The Chaplygin gas model merges dark matter and dark energy into a single fluid, described by the equation of state $(p=\frac{A}{\rho^{\alpha}})$ , which evolves from dust-like behavior in the early universe to a cosmological constant at later times. This dual function offers an appealing framework for understanding cosmic evolution. Although the model has faced challenges from observational data (as outlined below), its theoretical elegance continues to drive ongoing research, particularly with new datasets like those from DESI, which are prompting a revaluation of alternative dark energy models.\\

While the Chaplygin gas model holds considerable theoretical appeal, it has encountered significant challenges when compared to observational data. Multiple studies have demonstrated that the generalized Chaplygin gas (GCG) leads to oscillations and instabilities in the matter power spectrum, causing it to poorly match large-scale structure and Cosmic Microwave Background (CMB) observations \cite{sandvik2004, amendola2003}. In particular, the non-zero sound speed in GCG models suppresses the growth of matter perturbations, which contradicts the observed patterns of galaxy clustering and weak gravitational lensing. Further analysis, utilizing data from WMAP, Planck, and BAO surveys \cite{xu2014, lu2009}, has refined these constraints, leading many researchers to conclude that the simplest versions of the GCG are either ruled out or strongly disfavored \cite{bean2003}.\\

Recent findings from the Dark Energy Spectroscopic Instrument (DESI) have reignited interest in exploring non-traditional dark energy models, such as those involving Chaplygin-type fluids. However, it is crucial to recognize that this renewed focus is driven by particular observational anomalies and should not be interpreted as a general revival of the original Chaplygin gas models \cite{desi2023, alam2023}. While these anomalies currently remain within reasonable levels of tension, they have sparked theoretical investigations into alternative models that could potentially offer a better explanation for the accelerating expansion of the universe. Modified or extended Chaplygin gas models, for example, are now being considered as viable candidates for a unified dark energy sector that may account for these small discrepancies \cite{bernui2022, mamon2023}.\\

Despite the enthusiasm surrounding DESI’s findings, these conclusions have faced scrutiny. Several recent studies have critically assessed the assumptions, methods, and interpretations underlying the reported tensions with the standard \textbf{$\Lambda$CDM} model. Three recent studies critically assess the implications of the DESI 2024 dataset for various dark energy models, casting doubt on the standard \textbf{$\Lambda$CDM} model. The first study investigates different thermodynamic and phenomenological dark energy models and finds that a \textbf{$\Lambda$CDM} model, or a more complex log-corrected dark energy scenario, fits the DESI data better than the widely employed Chevallier-Polarski-Linder (CPL) parameterization, especially when considering the debated data point at $z=0.51$ \cite{paper1}. The second study uses the latest BAO data from DESI in combination with Hubble and supernovae measurements to derive model-independent constraints on the universe’s kinematics, including the deceleration, jerk, and snap parameters. It reports significant deviations in the jerk parameter, pointing to a potential tension with \textbf{$\Lambda$CDM}, although the Hubble tension is reduced when higher-order cosmographic terms are considered \cite{paper2}. The third study calibrates gamma-ray burst data by performing Bézier interpolation on the Hubble rate with the updated DESI data, finding that while \textbf{$\Lambda$CDM} remains the favored model, there is some weak statistical support for alternatives like non-flat or mildly evolving dark energy models, particularly when excluding the $z=0.51$ data point \cite{paper3}. Overall, these findings suggest that while the DESI data still align closely with the \textbf{$\Lambda$CDM} framework, the possibility of new dark energy dynamics or non-flat cosmologies remains open due to some subtle discrepancies. \cite{paper1} questions the statistical significance of DESI’s late-time Hubble parameter measurements, suggesting that potential systematic errors may not have been fully accounted for. In a similar vein, \cite{paper2} identifies possible issues with the calibration of galaxy redshift distributions and their subsequent effects on clustering analyses. Furthermore, \cite{paper3} points out the potential biases that may arise from the fiducial cosmological model used in DESI's simulation pipeline. These critical viewpoints must be carefully considered in any comprehensive evaluation of the implications of DESI’s results.\\

Recent theoretical advances have helped to resolve the long-standing issues associated with the Chaplygin gas model by introducing more refined variations that preserve its unifying potential while addressing its observational limitations. One particularly significant development \cite{MPS} presents a generalized framework that adjusts the equation of state by incorporating scale-dependent parameters. This modification effectively resolves the issue of the excessively high sound speed at intermediate redshifts and aligns the model's predictions more closely with observational data on both the background evolution and structure formation. Additionally, the refinement introduces an effective field-theory formulation, which not only strengthens the model's reliability but also allows for easier comparisons with a wider range of cosmological models. These improvements suggest that, while the original Chaplygin gas model may not be viable in its most basic form, carefully designed extensions offer a promising direction for future research, supported by both theoretical progress and observational evidence. \textbf{There has been a recent work \cite{DLM} of interest based on a generalized unified dark energy model involving a non-minimal interaction between a tachyonic fluid and a second scalar field with vacuum energy induced by a symmetry-breaking mechanism. This leads to an effective equation of state (EoS) that combines a generalized Chaplygin gas (GCG) component with a cosmological constant, which can be reinterpreted using the Murnaghan EoS from solid-state physics. The model exhibits a cosmological evolution that interpolates between a logotropic fluid at intermediate times, standard $\Lambda$CDM behavior at early times, and Chaplygin gas-like behavior at late times. Thermodynamically, it behaves like a compressible medium with non-zero pressure. Linear perturbation analysis has shown suppressed growth at early times and enhanced growth at late times, improving agreement with certain cosmological observations. Overall, the model offers a consistent and observationally viable alternative to $\Lambda$CDM, unifying dark matter and dark energy into a single matter-like fluid with non-zero pressure.}\\

The main component that we shall consider for our analysis is a dark fluid alternative to the GCG motivated from higher dimensional string theory \cite{Hova}. Then we shall consider coexistence of other alternative exotic matter components along with this fluid in the FU. The family of Chaplygin gas like fluids characterized by non-linear EoS can be derived from the Born-Infeld Lagrangian density and stands as a viable candidate for obtaining late-time acceleration besides providing a possible framework for unification of DE and dark matter(DM)\ cite{Bento,Debnath}. However, investigations of matter power spectrum neglecting baryon effects have ruled out GCG as a realistic candidate for unified DM and DE description \cite{x1,x2}. This motivates us to consider an alternative GCG model and study the effects of its coexistence with other fluids or fields following previous studies with modified Chaplygin gas \cite{ud1}. The existence of tachyons is predicted from both open and closed string theories indicating the instability of the perturbative vacuum. The tachyon field is believed to evolve towards a true vacuum characterized by zero energy density and this evolution can be described by an effective field theory as proposed by Sen \cite{Sen1}. The Chaplygin gas discussed earlier can also be arrived at from Superstring theories by considering a constant potential for the tachyon independent of the field in the effective formalism\cite{J}. The dilaton is another scalar field that find its origin in String theory from the lowest order low energy effective action that can result in averting the initial singularity problem in cosmology \cite{Brandenberger}. Both the tachyon and dilaton fields have been found to be useful as probable DE candidates \cite{Bagla,j2}. Another class of DE models arising from string theory is the DBI-essence which originates from the idea of dynamic branes in the extra dimensions, where the interbrane distance can act as a scalar field \cite{j3}. The effective scalar field model is characterized by a non-canonical kinetic term of the DBI form arising from the action obtained as a measure of the volume traversed by the dynamic brane, which turns out to be the given by square root of the induced metric.

In FU, the different cosmological
properties and quantum gravity phenomena has been discussed by Calcagni \cite{1GC10}, \cite{Calcagni}.
Also it has been investigated the fractal features of quantum gravity and D dimensional cosmology. The properties of a scalar field model have
also been investigated. Calcagni have
investigated the cosmology and Multi-scale gravity \cite{G. Calcagni}. Also the cosmic scenario in FU has been investigated by Das et al \cite{Dutta}. In the framework of FU new agegraphic, holographic and ghost DE models have been studied in \cite{K. Karami} ,\cite{Sadri} and interacting DE models with interaction between DE and DM  have been studied in \cite{Lemets}.
Thermodynamical properties in FU have analyzed by Sheykhi et al \cite{Teimoori}. Chattopadhyay et al and Jawad et al have been separately studied some special forms of holographic Ricci DE and pilgrim DE models in FU \cite{Roy}, \cite{Rani}. Also Halder et al and Debnath et al separately works on FU \cite{Haldar},\cite{Maity}. The idea of co-existing or correspondences or reconstruction procedure among different DE models become an interesting  unsolved theory  in cosmology. The idea  of correspondence
between various DE models
and their cosmological implications have been studied by several researchers 
[\cite{karami}–\cite{A. Pasqua}]

Motivating by all of these works on the FU, we have studied
the co-existence of alternative GCG and other DE in the Framework of FU. In this work, we have studied the alternative GCG in the FU in Sec II. In the subsections of this section, we have discussed the co-existence of DBI-essence, Techyonic field, Dilaton DE, Quintessence, k-essence, Hessence and alternative GCG models. Finally, we have written the conclusion of the whole work in Sec. III .

\section{\normalsize\bf{Alternative Generalized Chaplygin Gas in FU}}

The total action in fractal space-time is- \cite{1GC10}
\begin{equation}
S=S_{gr}+S_{mt}
\end{equation}
The gravitational action is given by,
\begin{equation}
S_{gr}=\frac{1}{2 M^{2}}\int
dQ(x)\sqrt{-g}(R-2\Lambda-\omega\partial_{\mu}v\partial^{\mu}v)
\end{equation}
and the matter action can be expressed as
\begin{equation}
S_{mt}=\int dQ(x)\sqrt{-g}{\cal L}_{m}
\end{equation}
Here $M^{2}=8\pi G$. We find the determinant value of the
dimensionless metric $g_{\mu\nu}$ denoted as $g$ and the cosmological constant $\Lambda$ is proportional to the fractal parameter $\omega$. Also, the term ${\cal L}_{m}$ denotes the Lagrangian matter, $v$ is a fractional function, and $dQ(x)$ is the Lebesgue-Stieltjes measure. Additionally, it is important to note that the measure $v$ is not a variable quantity like a scalar field or a dynamical object. Instead, its form is determined by the underlying geometry from the start.\\

In $D$-dimensional non-flat Friedmann-Robertson-Walker (FRW) universe the line-element takes the form
\begin{equation}
ds^{2}=-dt^{2}+a^{2}(t)\left[\frac{dr^{2}}{1-kr^{2}}+r^{2}d\Omega^{2}_{D-2}
\right]
\end{equation}
here $a(t)$ denotes the scale factor and $k~(=0,\pm 1)$ is the
curvature scalar.\\

Now in FU we evaluate the Friedmann equations as \cite{1GC10},
\begin{equation}
\left(\frac{D}{2}-1\right)H^{2}+\frac{k}{a^{2}}+H\frac{\dot{v}}{v}-\frac{1}{2}\frac{\omega}{D-1}\dot{v}^{2}
=\frac{M^{2}}{D-1}~\rho+\frac{\Lambda}{D-1}
\end{equation}
and
\begin{equation}
(D-2)\left(\dot{H}+H^{2}-H\frac{\dot{v}}{v}+\frac{\omega}{D-1}\dot{v}^{2}
\right)-\frac{\Box v}
{v}=-\frac{M^{2}}{D-1}~[(D-3)\rho+(D-1)p]+\frac{2\Lambda}{D-1}
\end{equation}
where the Hubble parameter $H=\frac{\dot{a}}{a}$ and $\Box v$ can be written as
\begin{equation}
\Box
v=\frac{1}{\sqrt{-g}}~\partial^{\mu}(\sqrt{-g}~\partial_{\mu}v)
\end{equation}
For simplicity we rewrite this equation as:
\begin{equation}
\Box v=-[\ddot{v}+(D-1)H\dot{v}]
\end{equation}
The continuity equation is
\begin{equation}
\dot{\rho}+\left[(D-1)H+\frac{\dot{v}}{v}\right](\rho+p)=0
\end{equation}
Assuming that the universe is filled with a dark fluid which may be considered as an 
alternative to (GCG), where the EoS is expressed as  \cite{Hova}
\begin{equation}
p=-\rho+\rho~sinc(\frac{\mu\pi\rho_{0}}{\rho})
\end{equation}
being $sinc(x) = \sin(x)/x$ and $\rho$ the dark fluid density,
which plays the role of the mixture of DE and DM densities.
This type of a phenomenological model is interesting as there is a transition from the deceleration epoch to an epoch of accelerated expansion with minor and simple modifications to the standard $\Lambda$ cold dark matter(\textbf{$\Lambda$CDM}) behavior. Just like GCG, the microscopic origin of this type of fluid can be explained using the extra dimensional theories. There is a supposed mixture of DE and DM. The EoS at high redshifts behaves as that of DM and at the present epoch the fluid can be described effectively with an EoS parameter close to -1. A huge advantage is the absence of future finite time singularities that plague most DE models. Moreover, the evidence of DE evolution can be well traced by the evolving EoS.
The dimensionless fine-tuning parameter $\mu$ appearing in the considered EoS can be constrained from stellar age bounds to be of the order of $\mu\gtrsim 0.688$ and the current value of fluid density can be constrained using the corresponding density parameter to be $\Omega_{\rho_{0}}\thicksim 0.96$ \cite{Hova}. Cosmological constraints on the model have been studied in recent times\cite{Almada}. We have used recent observations of SNe Ia and Hubble parameter measurements to constrain the fine-tuning parameter $\mu=0.843^{+0.014}_{-0.015}$. The present value of the deceleration parameter considering the above mentioned EoS takes the form $q_0=-0.67\pm0.02$ and the redshift at which the universe starts to dominated by the dark fluid preceded by the matter-dominated era turns out to be given by $z_{t}=0.57\pm 0.04$.

Using the equations (8) and (9), we get
\begin{equation}
\rho=\frac{\mu\pi\rho_{0}}{2\tan^{-1}(a^{\alpha+D-1}tan(\frac{\mu\pi}{2}))}
\end{equation}

As we know the time dependent fractional function $v$ is arbitrary,
we take the form as  $v=v_{0}a^{\alpha}$, where the constant $v_{0}$
and $\beta$ are positive.\\

From the equation (5) and (11) we get

\begin{eqnarray*}
H^{2}=\frac{\frac{M^{2}}{D-1}[\frac{\mu\pi\rho_{0}}{2\tan^{-1}(a^{\alpha+D-1}
\tan(\frac{\mu\pi}{2}))}]+\frac{\Lambda}{D-1}-\frac{k}{a^{2}}}{\frac{D}{2}-1+\alpha-
\frac{\omega v_{0}^{2}\alpha^{2}a^{2\alpha}}{2(D-1)}}
\end{eqnarray*}
\begin{equation}
     =\frac{M^{2}\mu\pi\rho_{0}a^{2}+2(\Lambda a^{2}-k(D-1))(D-1)\tan^{-1}(a^{\alpha+D-1}
\tan(\frac{\mu\pi}{2}))}{\{ (D-1)(D+2(\alpha-1))-\omega v_{0}^{2}\alpha^{2}a^{2\alpha}\}a^{2}}
\end{equation}

\begin{figure}[thbp]
\centering
\includegraphics[width=0.35\textwidth]{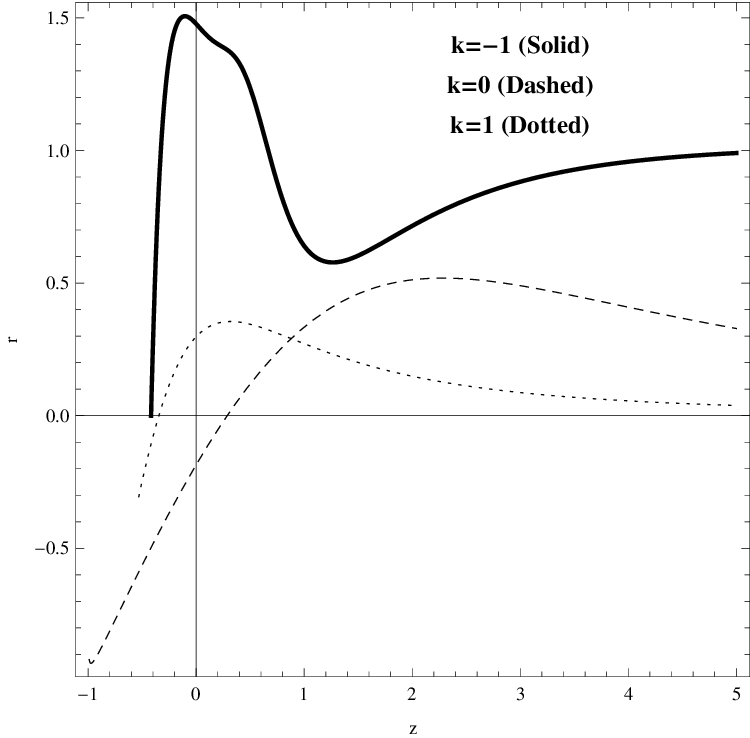}
\caption{Variation of $r$ with $z$ for GCG in FU} \label{fig1}
\end{figure}

\begin{figure}
\includegraphics[width=0.35\textwidth]{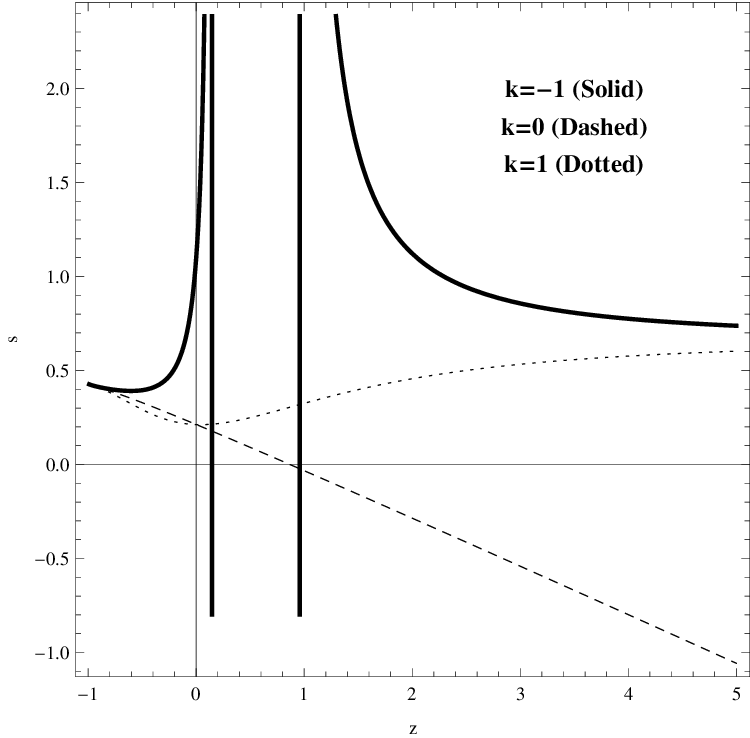}~~~~~~~~
\includegraphics[width=0.35\textwidth]{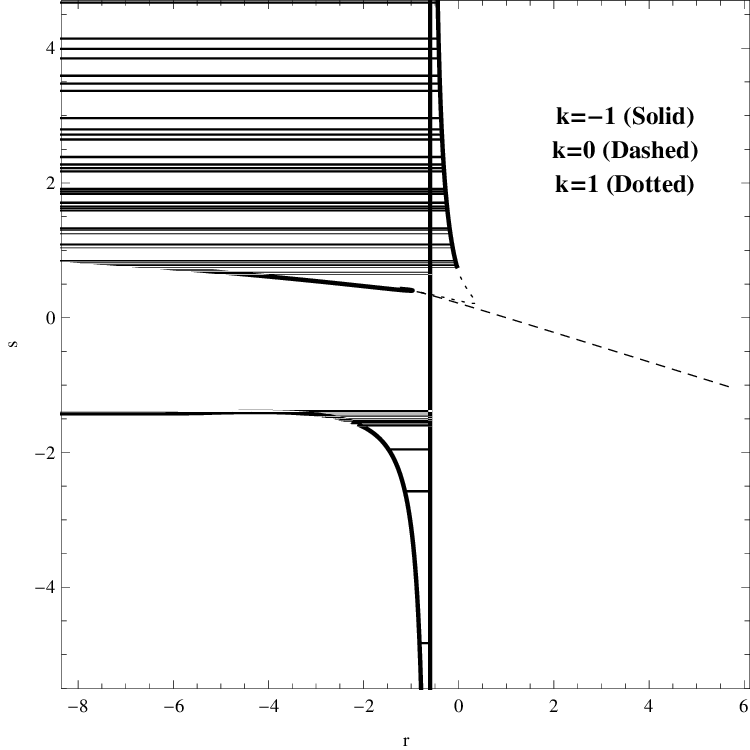}\\
Fig.2:Variation of $s$ with $z$ for GCG in FU~~~~~~~Fig.3:Variation of $r$ with $s$ for GCG in FU~~\\
\vspace{2mm}

\includegraphics[width=0.35\textwidth]{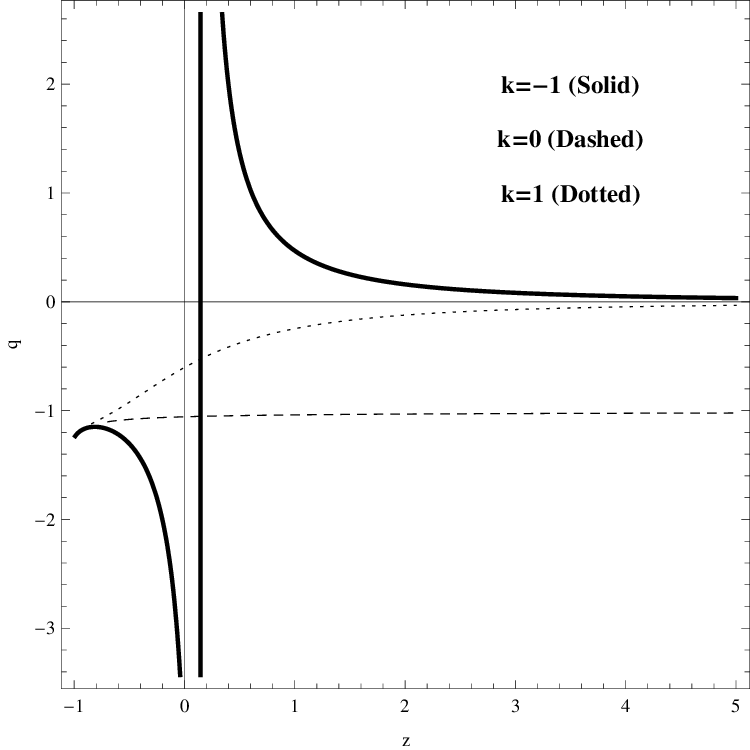}~~~~~~
\includegraphics[width=0.35\textwidth]{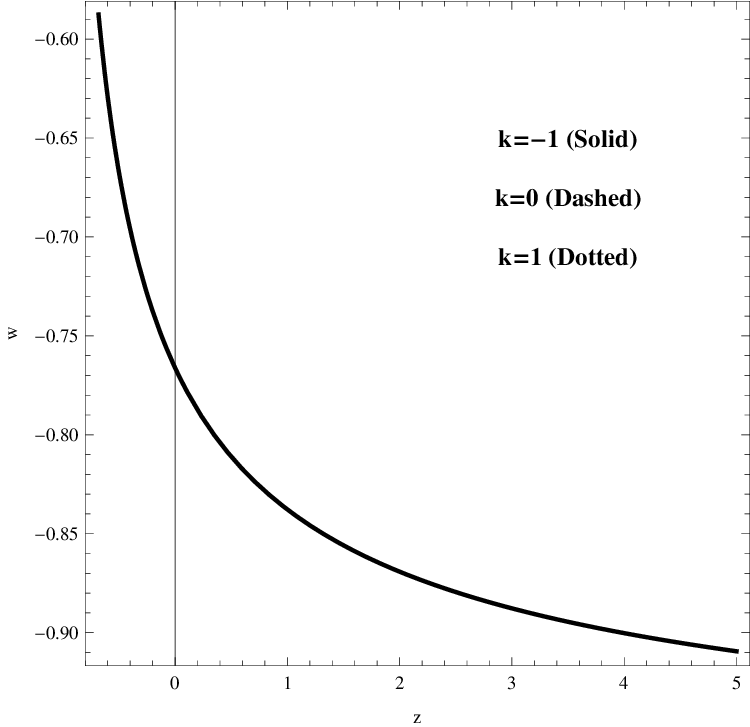}\\
~~~Fig.4:Variation of $q$ with $z$ for GCG in FU~~~~~~~Fig.5:Variation of $w$ with $z$ for GCG in FU~~\\

\end{figure}

The equation of state parameter w can be written as,
\begin{equation}
w=-1+sinc(2\tan^{-1}(a^{\alpha+D-1}\tan(\frac{\mu\pi}{2})))
\end{equation}
The deceleration parameter is
\begin{equation}
q=-1-\frac{\dot{H}}{H^{2}}
\end{equation}
The form of statefinder diagnostics $\{r,s\}$ are
\begin{equation}
r=\frac{\dddot{a}}{aH^{3}}=q\left(1-\frac{2}{aH^2}\right)-\frac{a^{2}}{H^{2}}\left(\frac{dH}{da}\right)^{2}
-(1+q)+\frac{a^{2}}{H}\frac{d^{2}H}{da^{2}}
\end{equation}
and
\begin{equation}
s=\frac{r-1}{3(q-\frac{1}{2})}
\end{equation}
As we know $H^{2}$ from eq. (12), whose scale factor term is $a$ i.e., in term of the
redshift $z~(=\frac{1}{a}-1)$. To get the expressions of $q$, $r$ and
$s$ in term of $z$ we put the expression of $H^{2}$ in equations (14),
(15) and (16). But we will get very lengthy expressions which are not
indicate in this work. For this reason, in graphically we draw the
natures of the parameters. The value of $q$, $r$ and
$s$ calculated by using Mathematica software and the figure are drawn numerically. From figure 1, we get an open universe,
i.e., $r$ decreases when $z$ decreases and after a certain time $r$
increases when $z$ decreases for the value of $k=-1$. Also from  the figure 1,
we find the $r-z$ relation for a universe which is closed due to $k=1$,
$r$ decreases when $z$ decreases. From figure 2, 
we find the relations between $s$ and $z$ for closed and open universe
respectively. The parameter of state-finder $s$ at first decreases when
$z$ decreases and after some time it gradually increases when $z$ decreases
for a closed universe. In case of open universe, the peculiar results
of $s$ happen as $z$ decreases. In figure 3 the $s$ vs $r$
are shown for closed and open universe respectively. For closed and open
universe, we find the relations between $q$ and $z$ in the figure 4. Both of the closed and open universe we find out that
the MCG crosses the phantom divides when $z$ decreases. $w$ vs $z$ for closed
and open universe  are drawn in the figure of 5.
For both of the closed and open universe, the equation of state parameter
$z$ decreases when $w$ increases.\\

In the next subsections, we will discuss the co-existence of DBI-essence, Techyonic field, Dilaton DE,
Quintessence, k-essence, Hessence and alternative GCG in FU.

\subsection{\normalsize\bf{Co-existence of DBI-essence and alternative GCG in FU}}

The idea of DBI-essence was motivated from the probable connection between the superstring /M- theory and inflation which is also supposed to be a phase of rapid accelerated expansion in the very early universe that can seed the necessary cosmological perturbations leading to the formation of large scale structures later on in the universe. In standard General Relativistic cosmology, the scalar field potentials required to generate an inflationary and late-time accelerating phase are quite different in nature, as the former requires a flat potential due to rapid exponential expansion while the later requires a steep potential due to much milder acceleration rate. However, in the extra dimensional scenario involving strings and branes, this drastic difference in the nature of the two potential is not necessarily true \cite{Kach1,Kach2}. The energy density and pressure of the scalar field in this case may be expressed in the form \cite{gg}

\begin{figure}
	\includegraphics[width=0.35\textwidth]{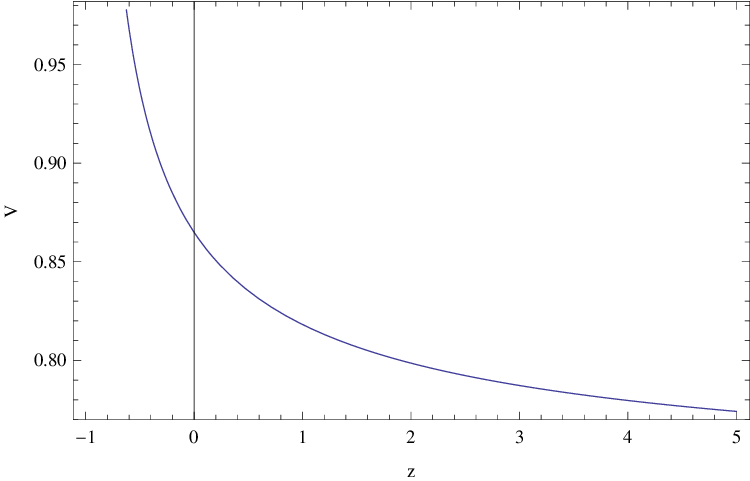}~~~~~~~~
	\includegraphics[width=0.35\textwidth]{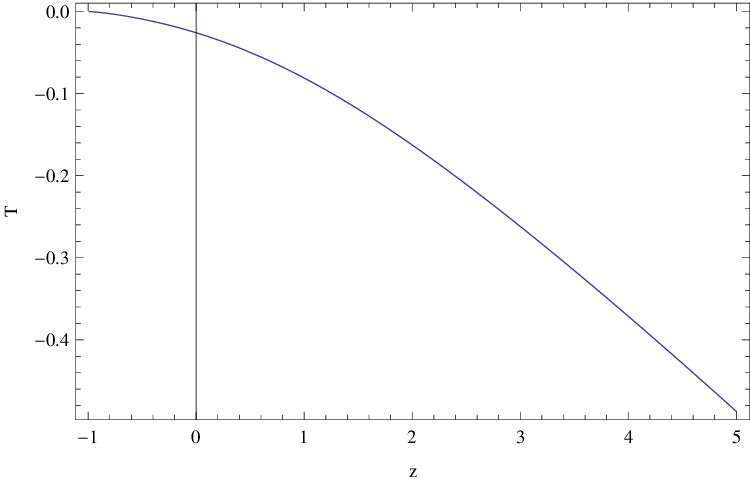}\\
	Fig.6(a):variations of $V$ against $z$ for DBI-essence in FU~~~Fig.6(b):variations of $T$against $z$ DBI-essence in FU~~\\
	\vspace{2mm}
	
	\includegraphics[width=0.35\textwidth]{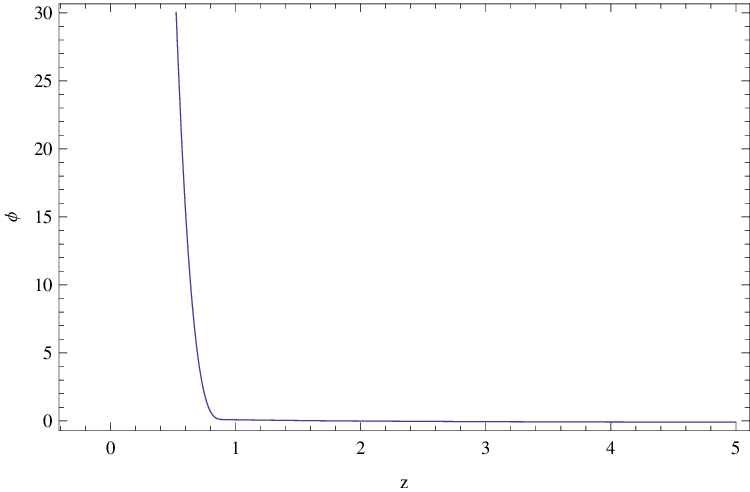}~~~~~~
	\includegraphics[width=0.55\textwidth]{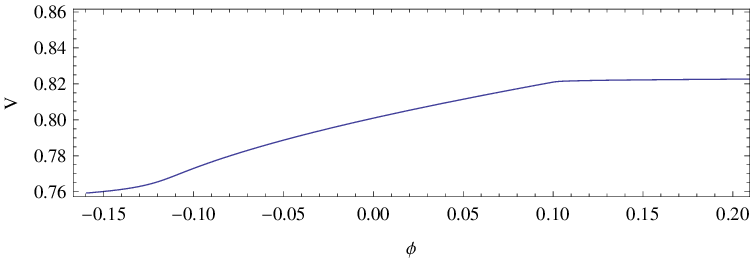}\\
	~~~Fig.6(c):variations of $\phi$ against $z$ DBI-essence in FU~~~~~~~~~~~Fig.6(d):variation of
	$V$ against $\phi$ DBI-essence in FU~~\\
	
\end{figure}

\begin{equation}
\rho_{d}=(\delta-1)T(\phi)+V(\phi)
\end{equation}
and
\begin{equation}
p_{d}=\frac{\delta-1}{\delta}T(\phi)-V(\phi)
\end{equation}
where $\delta$ is given by
\begin{equation}
\delta=\frac{1}{\sqrt{1-\frac{\dot{\phi}^{2}}{T(\phi)}}}
\end{equation}
where $V(\phi)$ is the self interacting potential and $T(\phi)$ is
the warped brane tension.\\
Considering co-existence of DBI-essence and alternative GCG fluid considered in the background of the FU, we take $\rho_{D}=\rho$ and $p_{D}=p$ from (10), (11),
(17) and (18), we get
\begin{equation}
V(\phi)= \frac{\mu\pi\rho_{0}\left[1-\delta~sinc(2\tan^{-1}(a^{\alpha+D-1}
\tan(\frac{\mu\pi}{2})))\right]}{2(1+\delta)\tan^{-1}(a^{\alpha+D-1}
\tan(\frac{\mu\pi}{2}))},
\end{equation}
\begin{equation}
T(\phi)=\frac{\delta\mu\pi\rho_{0}~sinc(2\tan^{-1}(a^{\alpha+D-1}
\tan(\frac{\mu\pi}{2})))}{2(\delta^{2}-1)\tan^{-1}(a^{\alpha+D-1}
\tan(\frac{\mu\pi}{2}))}
\end{equation}
and
\begin{equation}
\phi=\int \frac{1}{aH}\sqrt{\frac{(\rho + p)}{\delta}}da,
\end{equation}
with the sum of energy density and pressure turns out to have the form
\begin{equation}
\rho + p=\frac{\mu\pi\rho_{0}~sinc(2\tan^{-1}(a^{\alpha+D-1}
\tan(\frac{\mu\pi}{2})))}{2\tan^{-1}(a^{\alpha+D-1}
\tan(\frac{\mu\pi}{2}))}
\end{equation}

The variations of $V$, $T$ and $\phi$ along with the redshift $z$ have been plotted in Figure 6. We also plot the variation of $V$ against the scalar field $\phi$. It is found that both $V$ and $T$ decreases as we go to higher redshifts. However, $V$ and $T$ keeps increasing in the immediate future just like the scalar field $\phi$ indicating that the accelerating expansion will continue in the recent future. A plot of the scalar field potential against the field shows a steady increase before the potential becomes almost flat and there is no significant variation along the field.

\subsection{\normalsize\bf{Co-existence of Tachyonic field and alternative GCG in FU}}

The idea of tachyon matter is deep rooted in the extra dimensional superstring theories\cite{Sen1,Sen}. Tachyon matter has been considered to be a probable DM candidate \cite{Davies}. Such a consideration may be verified from modified temperature-time relation of the cosmic constituents or an excess in particle production in the early epoch arising from quantum corrections. Tachyonic fields have also found application in literature as probable DE candidates \cite{Srivastava,Bagla,Avelino}. The late time acceleration can be accounted for by using a self interacting non-minimally coupled tachyon field described by an inverse cubic potential with subsequent decay to cold DM leading to a probable resolution of the coincidence problem \cite{Srivastava}. An alternate approach considering coexistence of tachyon matter with non-relativistic matter as well as radiation yields a tachyon density comparable to matter even at higher redshifts corresponding to the matter dominated epoch\ cite{Bagla}. It has been found that for a tachyon field described by an exponential potential, the late-time acceleration is again followed by an epoch of matter domination, thus eliminating the possibility of the acceleration leading to future finite-time singularities, a problem plaguing most DE models.  Avelino et al.~\cite{Avelino} found a similar evolution of the Hubble parameter and the field potential for tachyon and quintessence fields but the field itself evolves quite differently as a function of redshift in both the cases. 

\begin{figure}
	\includegraphics[width=0.35\textwidth]{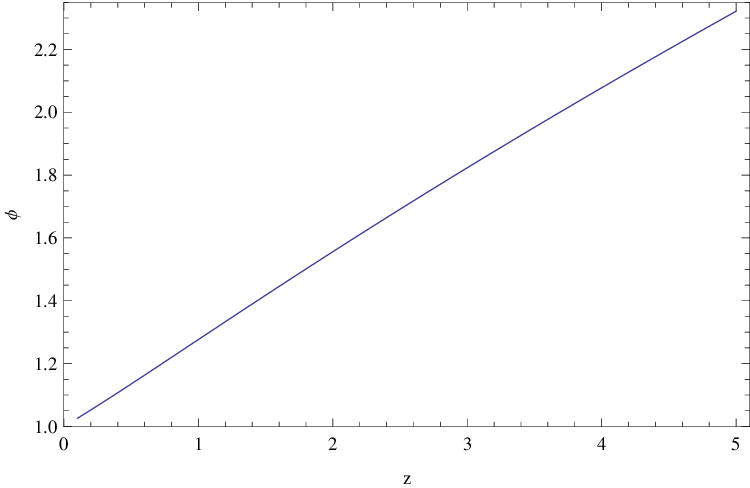}~~~~~~~~
	\includegraphics[width=0.35\textwidth]{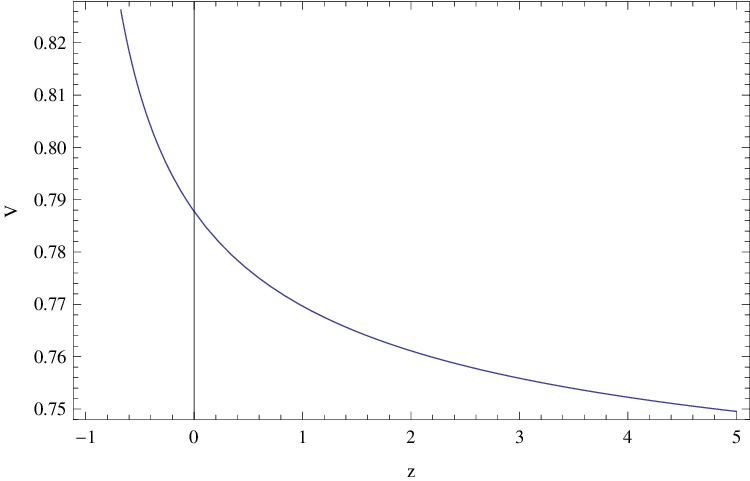}\\
	
	~~~~Fig.7(a):Variations of $\phi$ against $z$ for Tachyonic field in FU~~~~~~~~~Fig.7(b):Variations of $V$ against $z$ for Tachyonic field in FU~~\\
	
	\vspace{3mm}
	
	\includegraphics[height=2.0in]{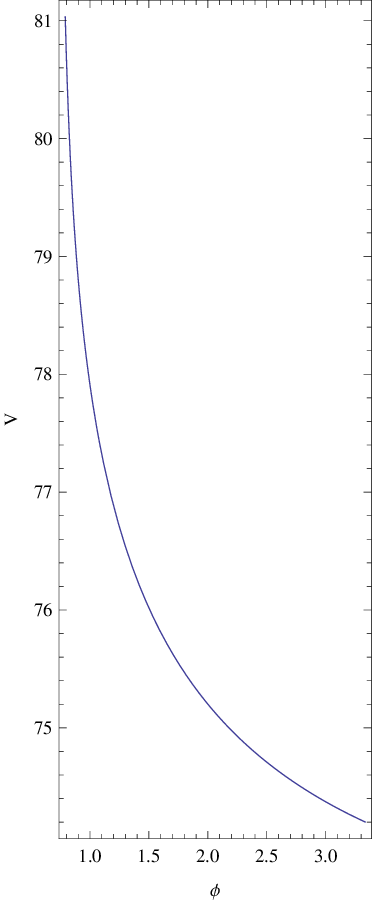}\\
	
	~~~~~~~~Fig.7(c):variation of $V(\phi)$
	against $\phi$ for Tachyonic field in FU~~~~~~~~~~~~~~~~~~~
\end{figure}

The energy density and pressure of the tachyon field can be expressed in the form

\begin{equation}
\rho_{t}=\frac{V(\phi)}{\sqrt{1- \dot{\phi}^{2}}}
\end{equation}
and
\begin{equation}
p_{t}=-V(\phi)\sqrt{1- \dot{\phi}^{2}}
\end{equation}
where $\phi$ is the tachyonic field and $V(\phi)$ is the
corresponding potential. Using (10), (11), (24) and (25) we obtain the scalar field to be given by
\begin{equation}
\phi=\int \frac{\sqrt{1+\omega}}{aH}da
\end{equation}
and the corresponding potential to be given by
\begin{equation}
V=\sqrt{-\rho_{t}p_{t}}
\end{equation}
where
$\omega=-1+~sinc(2\tan^{-1}(a^{\alpha+D-1}\tan(\frac{\mu\pi}{2})))$.

The variations of the field against the redshift has been plotted in Fig. 7(a). The evolution of the field against the redshift is found to be almost linear in nature. However, the variation of the corresponding potential against the redshift is found to be gradually decaying as we move to higher redshifts indicating a possible transition to the matter dominated epoch. The variation of the field potential against the tachyon field that we have plotted in Figure 7(c) is found to have a steep decreasing nature as desired for explaining the current DE dominated era. On the contrary, a flat potential is used to describe tachyonic inflation\ cite{RPS,Li}.

\subsection{\bf{Co-existence of dilaton DE and alternative GCG in FU}}

The dilaton field also originates from the extra dimensional Superstring theories and is associated with the varying volume of the compactified higher dimensions. It can be  manifested in the form of a radion field modulating the inter-brane distance in warped braneworld models. The possibility of the dilaton field as a possible DE candidate has been explored in the framework of Weyl-Scaled induced gravity\cite{Lu}. An exponential dilatonic potential has been found capable of causing the universe to accelerate at late-times with a quicker growth of structure rate than the \textbf{$\Lambda$CDM} and cannot be ruled incompatible with the CMBR data.

The energy density and pressure of the dilaton field has the form \cite{j2}

\begin{equation}
	\rho_{D}=-\chi+3\zeta e^{\lambda \phi}\chi^{2}
\end{equation}
and
\begin{equation}
	p_{D}=-\chi+\zeta e^{\lambda \phi}\chi^{2},
\end{equation}

where $\chi=\frac{1}{2}\dot{\phi}^2$ denotes the kinetic term, the parameter $\lambda$ has the dimension of length and modulates the interactions of the dilaton field which are of non-gravitational nature and $\zeta$ is an arbitrary constant that must be positive.

Following the same prescription of comparing the energy density and pressure of alternative GCG in FU and the co-existing scalar field as used in the previous cases, the dilaton field turns out to be given by

\begin{equation}
	\phi=\frac{4}{\lambda} \log
	\left[\frac{\lambda}{4}\left(\frac{2}{\zeta}\right)^{^{\frac{1}{4}}}\int
	\frac{1}{a H}\left(\rho-p\right)^{\frac{1}{4}}da\right]
\end{equation}

where
\begin{equation}
	\rho - p=\frac{\mu\pi\rho_{0}\{2-~sinc(2\tan^{-1}(a^{\alpha+D-1}
		\tan(\frac{\mu\pi}{2})))\}^{2}}{2\tan^{-1}(a^{\alpha+D-1}.
		\tan(\frac{\mu\pi}{2}))}
\end{equation}

In figure 8 we have drawn $\phi$ against $z$ and find the evolution of the field with the redshift shows that $\phi$ decreases as $z$ decreases.

\begin{figure}
	\includegraphics[height=2.0in]{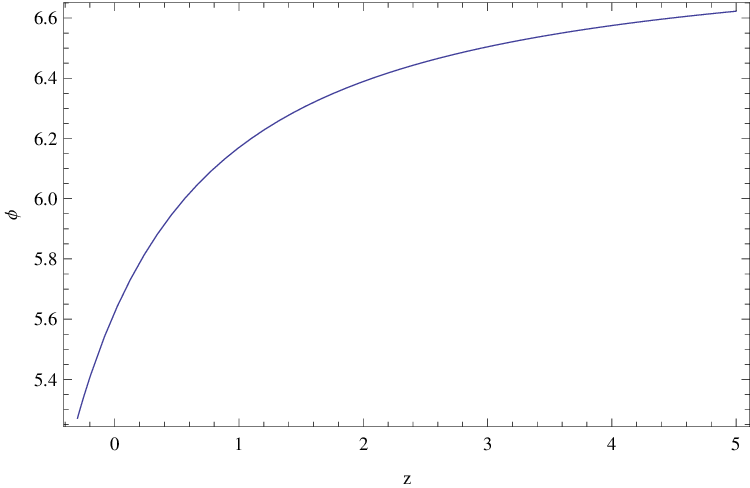}~~~\\
	
	~~~~~~~~Fig.8:Variation of $\phi$ against $z$ dilaton DE in FU~~~~~~~~~~~~~~\\
\end{figure}

\subsection{\bf{Co-existence of Quintessence and alternative GCG in FU}}

The idea of quintessence\cite{q1,q2} involves a canonical scalar field to account for the present cosmic acceleration. In this case the equation of state parameter for the DE is no longer a constant but evolves with time. For the class of quintessence models described by scalar field potentials characterized by tracking or thawing properties, an analytical expression can be obtained for the varying EoS parameter involving a number of model parameters. The analytically obtained EoS parameter can be used to constrain the corresponding quintessence model from different data of the observations like type SNe Ia, CMB or BAO. From the constraints it appear that the tracker models are in close approximately with the standard \textbf{$\Lambda$CDM}, but the thawing class of potentials can yield observational constrained quintessence models with a different value of the present field EoS parameter.

The energy density and pressure corresponding to the quintessence field can be expressed in the form

\begin{equation}
	\rho_{Q}=\frac{1}{2}\dot{\phi}^{2}+V(\phi)
\end{equation}
and
\begin{equation}
	p_{Q}=\frac{1}{2}\dot{\phi}^{2}-V(\phi)
\end{equation}

The scalar field turns out to be expressed by the form
\begin{equation}
	\phi=\int \frac{1}{aH}(\rho+p)^{\frac{1}{2}}da
\end{equation}

The corresponding potential is given by

\begin{eqnarray*}
	V(\phi)= \frac{1}{2}(\rho-p)
\end{eqnarray*}
\begin{equation}
	=\frac{\mu\pi\rho_{0}\{2-~sinc(2\tan^{-1}(a^{\alpha+D-1}
		\tan(\frac{\mu\pi}{2})))\}^{2}}{4\tan^{-1}(a^{\alpha+D-1}
		\tan(\frac{\mu\pi}{2}))}
\end{equation}

\begin{figure}
	\includegraphics[height=2.0in]{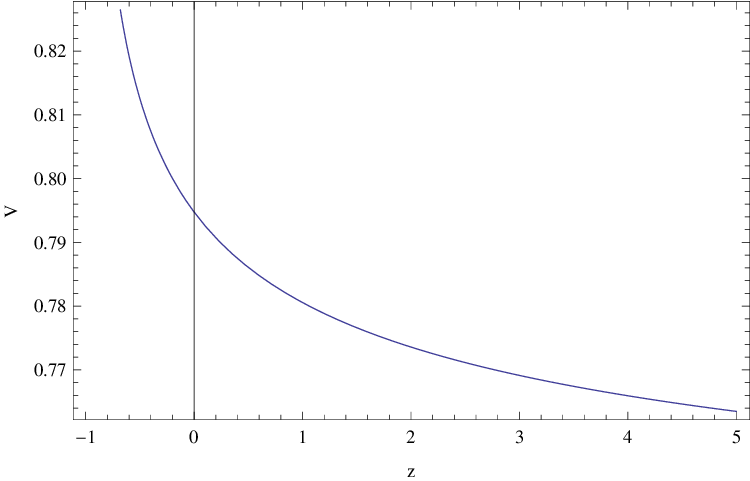}~~~~~~~~
	\includegraphics[height=2.0in]{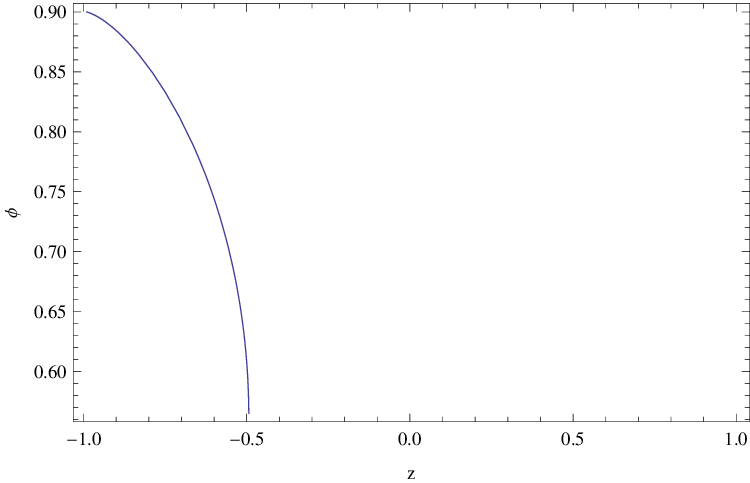}\\
	~~~~Fig.9(a):Variations of $V(\phi)$ against $z$ for Quintessence in FU~~~~~~Fig.9(b):Variations of $\phi$ against $z$ for Quintessence in FU~~\\
	\vspace{2mm}
	\includegraphics[height=3.0in]{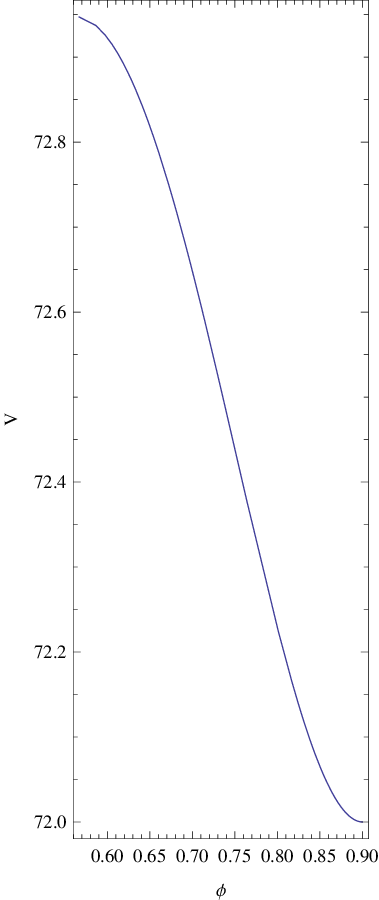}\\
	~~~~~~~~~~~~~~~~~~Fig.9(c):Variations of $V(\phi)$ against $\phi$ for Quintessence in FU~~~~~~~~~~~~~~~~~\\
	\vspace{3mm}
	
\end{figure}

In Figures 9(a) and 9(c), the obtained scalar field potential is varied against the redshift and the scalar field while in Fig. 9(b) we plot the variation of the field against the redshift. It is found that as the redshift increases, both the scalar field and its corresponding potential drops down significantly. The field experiences a sharper and more abrupt drop with increasing redshift. Both the variations suggest the acceleration to continue in the immediate future as their value keeps increasing for negative redshifts. The variation of the field potential with increasing field shows a steady and quick decline for short range of field values implying a steep potential as desirable for DE models.

\subsection{\normalsize\bf{Co-existence of k-essence and alternative GCG in FU}}

The k-essence models of DE were introduced mainly to avoid any fine-tuned model parameters or anthromic arguments. The model comprises of a dynamical attractor solution behaving as a $\Lambda$-term only as the matter-dominated epoch begins. The k-essence density then gradually grows and exceeds the matter density to induce the presently observed epoch of accelerated expansion. The k-essence scalar field evolves with a kinetic energy term in the action which is not linear in nature and the models are open to both possibilities of eternally continuing acceleration and acceleration discontinuing after finite time. The stress-energy components of the k-essence field may be expressed in the form \cite{k1,k2}
\begin{equation}
	\rho_{K}=V(\phi)(-\psi+3\psi^{2})
\end{equation}
and
\begin{equation}
	p_{K}=V(\phi)(-\psi+\psi^{2})
\end{equation}
where $\phi$ is the scalar field having kinetic energy $\psi=
\frac{1}{2}\dot{\phi}^{2}$ and $V(\phi)$ is
the potential of the k-essence field.\\

We compare the energy density and pressure of the alternative GCG fluid coexisting in the background FU with that of the k-essence scalar to compute the field and the potential, respectively, which turns out to have the form

\begin{equation}
	\phi=\int
	\frac{\sqrt{2}}{a H}\left[\frac{2-~sinc(2\tan^{-1}(a^{\alpha+D-1}
		\tan(\frac{\mu\pi}{2})))}{4-3~sinc(2\tan^{-1}(a^{\alpha+D-1}
		\tan(\frac{\mu\pi}{2})))}\right]^{\frac{1}{2}}da
\end{equation}
and
\begin{eqnarray*}
	V(\phi)=\frac{(\rho - 3p)^{2}}{2(\rho - p)}
\end{eqnarray*}
\begin{equation}
	=\frac{\mu\pi\rho_{0}\{4-3~sinc(2\tan^{-1}(a^{\alpha+D-1}
		\tan(\frac{\mu\pi}{2})))\}^{2}}{2\{2-~sinc(2\tan^{-1}(a^{\alpha+D-1}
		\tan(\frac{\mu\pi}{2})))\}}
\end{equation}

\begin{figure}
	\includegraphics[height=2.0in]{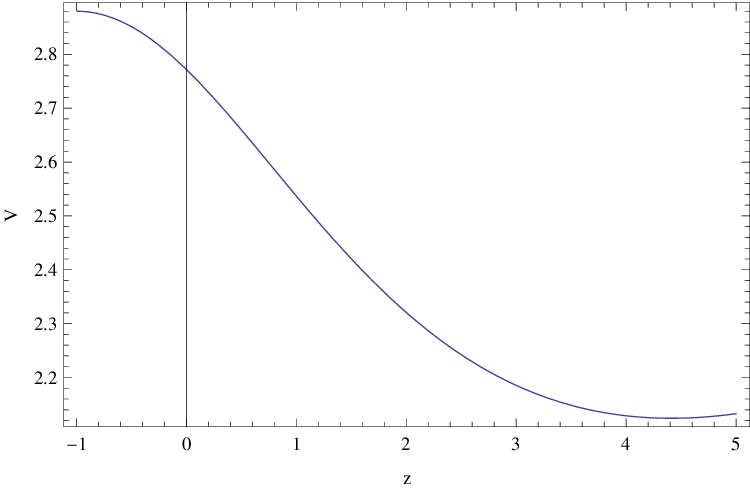}~~~~~~~~
	\includegraphics[height=2.0in]{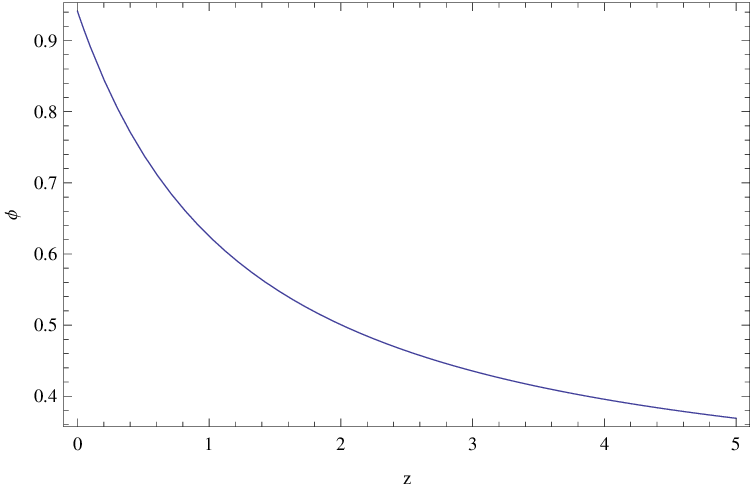}\\
	
	~~~~Fig.10(a): Variations of $V(\phi)$ against $z$ for k-essence in FU~~~~~~~Fig.10(b): Variations of $\phi$ against $z$ for k-essence in FU ~~\\
	\vspace{3mm}
	
	\includegraphics[height=2.0in]{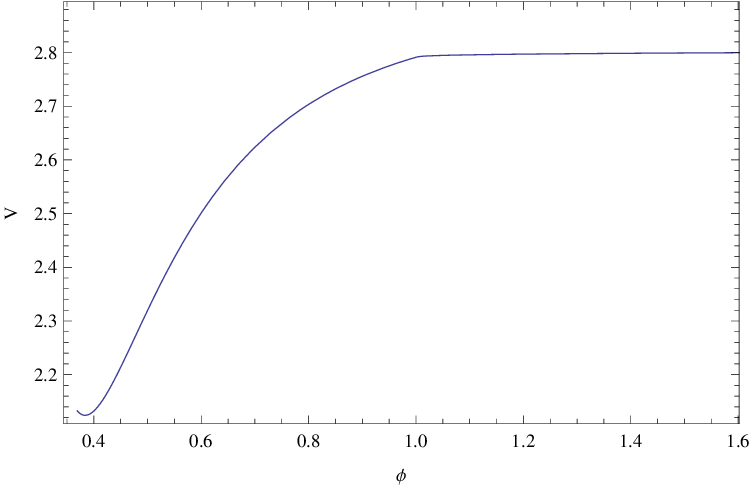}\\
	
	~~~~Fig.10(c): Variations of $V(\phi)$ against $\phi$ for k-essence in FU~~~~\\
	\vspace{3mm}
\end{figure}

The variation of the obtained potential and the field along the redshift have been shown in Figures 10(a) and 10(b), while the variation of the potential along the field has been shown in Figure 10(c). As we can see, the potential decreases as we go to higher redshifts just like the field does. The potential increases steadily as the field value increases but for large field values there is no significant increase in the potential.

\subsection{\normalsize\bf{Co-existence of Hessence and alternative GCG in FU}}

Observational data, including those from the Planck 2018 results and combined probes from BAO, SNe Ia, and the CMB, slightly favor a DE  fluid with an EoS parameter less than $-1$ \cite{Planck2018, DiValentino2021}. One such attempt to model this behavior is the quintom scenario, which allows the EoS to cross the phantom divide ($\omega = -1$) but does so at the expense of introducing two real scalar fields with opposite kinetic terms \cite{Feng2005, Cai2010}. This problem can be overcome in the hessence DE model which considers a single complex scalar field with an internal degree of freedom and is non-canonical. At late times it behaves like a Chaplygin gas on appropriate choice of potential.

The energy density and pressure of the hessence field has the form\cite{wei}

\begin{equation}
p_{H}=\frac{1}{2}(\dot{\phi}^{2}-\phi^{2}\dot{\vartheta}^{2})-V(\phi)
\end{equation}
and
\begin{equation}
\rho_{H}=\frac{1}{2}(\dot{\phi}^{2}-\phi^{2}\dot{\vartheta}^{2})+V(\phi),
\end{equation}
where ($\phi$, $\vartheta$) are the variables introduced to describe the hessence.

The total conserved charge $Q$ has the form
\begin{equation}
Q=a ^{3}\phi^{2}\dot{\vartheta}= \text{constant}.
\end{equation}

We apply the same technique as in the previous cases to obtain expressions for the field and corresponding potential, which turn out to have the form

\begin{equation}
\phi=\int
\frac{1}{aH}(\rho+p+\frac{Q^{2}}{a^{6}\phi^{2}})^{\frac{1}{2}}da
\end{equation}
and
\begin{equation}
V(\phi)=\frac{\mu\pi\rho_{0}[2-~sinc(2\tan^{-1}(a^{\alpha+D-1}
\tan(\frac{\mu\pi}{2})))]}{4\tan^{-1}(a^{\alpha+D-1}
\tan(\frac{\mu\pi}{2}))}
\end{equation}

The variation of the field and potential against the redshift have been plotted in Figs. 11 (a) and (b), and Fig 11(c) shows the variation of the potential corresponding to change in field values. The field is found to increase as the redshift increases suggesting coexistence of the field even during matter dominated epoch as expected for hessence. The potential however decreases for increasing redshift as well as field values.

\begin{figure}
\includegraphics[height=2.0in]{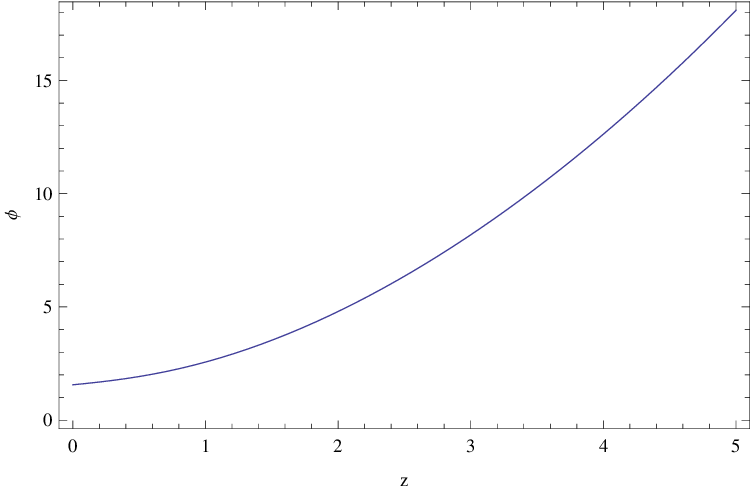}~~~~~~~~
\includegraphics[height=2.0in]{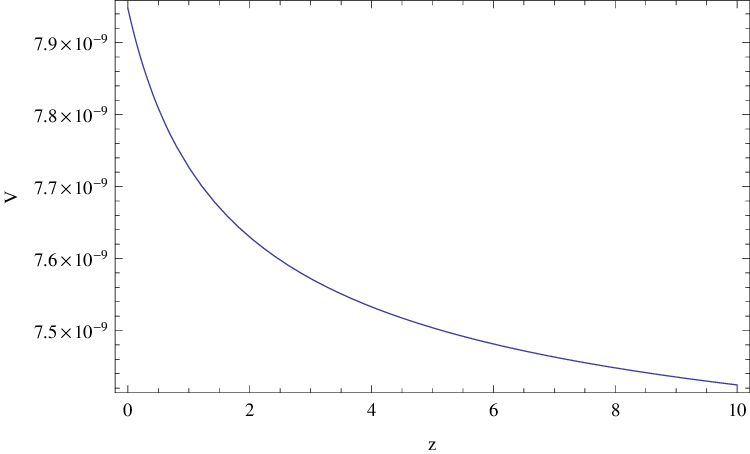}\\

~~~~Fig.11(a):Variations of $\phi$ against
$z$ for Hessence in FU~~~~~~~~~~~Fig.11(b): Variations of$V(\phi)$ against $z$ for Hessence in FU~~\\

\vspace{3mm}
\includegraphics[height=2.0in]{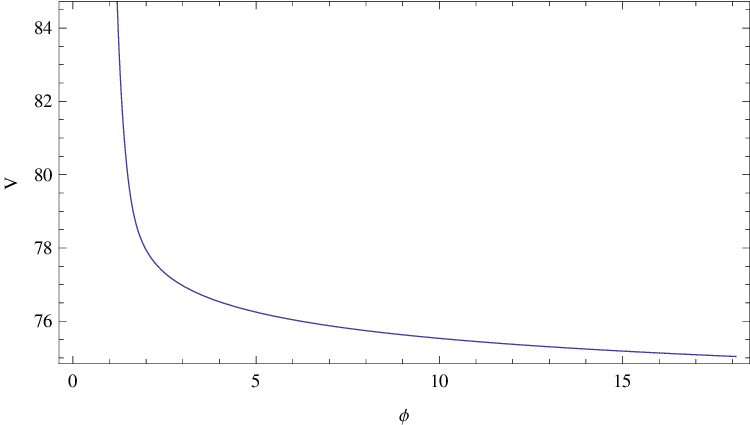}\\

~~~~~~~~~~~~~~~~~~~~~~~~Fig.11(c):Variations of$V(\phi)$ against $\phi$ for Hessence in FU~~~~~~~\\

\end{figure}

\section{\normalsize\bf{Conclusions}}

In this paper we have considered a fractal cosmological model in D-dimensions described by the FRW spacetime metric and we have assumed first that the evolution of the universe is dominated by an exotic fluid in the form of alternative GCG. Additionally, we have assumed that the fractional function obeys a power-aw form. The relevant cosmological parameters like the deceleration parameter, EoS parameter and the statefinder parameters have been computed and their variations with the redshift have been plotted in this setup for all three possible values of the spatial curvature $k=-1, 0, 1$. Also, the variation between the two statefinder parameters have been plotted. The co-existence of different DE to explain the present accelerating phase of the universe has been explored widely by Setare \cite{S1,S2,S3,S4,S5}. Following the same approach, we have extended our analysis by considering coexistence of alternative GCG and scalar fields originating from the extra dimensional quantum gravity theory like M-theory like tachyon, dilaton and DBI-essence as well as field theory motivated scalar fields like quintessence, k-essence and hessence. In each case, the evolution of the scalar field and potential has been obtained and their variation with the redshift have been plotted. In most of the cases the accelerated expansion of the universe appears to be continuing in the recent future. The variation of the potential along the field values has also been obtained. As we have discussed, we find that the coexistence of alternative GCG with these scalar fields originating from both higher dimensional and four dimensional physics in the background of a FU can be used to describe the current accelerating epoch of the universe to good effect. The entire investigation deviates from the standard big bang model as a whole including the corresponding \textbf{$\Lambda$CDM} description for the late-time acceleration in the sense that the usual isotropic and homogeneous background is replaced by the fractal spacetime structure and the cosmological constant plagued by different theoretical inconsistencies is replaced by a co-existing setup of an alternative GCG like fluid with different individual scalar fields at a time originating from both four and higher dimensional theories. Most of the resultant models are theoretically consistent and their observational viability shall be put to scrutiny in recent future works.\\

\end{document}